\documentclass[useAMS,usenatbib,letterpaper]{mn2e}

\usepackage{graphicx}
\usepackage{natbib}
\usepackage{aas_macros}
\begin{document}

\title[Quadruple system VW~LMi]{VW~LMi: tightest quadruple system known. Light-time effect 
       and possible secular changes of orbits \thanks{Based on the data obtained at the David Dunlap 
       Observatory, University of Toronto}}

\author[T. Pribulla et al.]{T. Pribulla$^{1,2}$, D. Balu\v{d}ansk\'y$^3$, P. Dubovsk\'{y}$^4$,
        I. Kudzej$^4$, \v{S.} Parimucha$^{5}$, M. Siwak$^6$, M. Va\v{n}ko$^{7,2}$\\
$^1$Department of Astronomy, University of Toronto, 50 St.~George St., Toronto, 
Ontario, Canada M5S~3H4\\
$^2$Astronomical Institute of the Slovak Academy of Sciences, 059\,60 Tatransk\'{a} 
Lomnica, The Slovak Republic,\\ E-mail: (pribulla,vanko)@ta3.sk\\
$^3$Roztoky Observatory, 090 01 Vy\v{s}n\'y Orl\'{\i}k, The Slovak Republic, E-mail: bdaniel@pobox.sk\\
$^4$Vihorlat Observatory, Mierov\'a 4, Humenn\'e, The Slovak Republic, 
E-mail: var@astro.sk, vihorlatobs1@stonline.sk\\
$^5$Institute of Physics, Faculty of Natural Sciences, University of P.J. \v{S}af\'{a}rik, 
040 01 Ko\v{s}ice, The Slovak Republic,\\E-mail: stefan.parimucha@upjs.sk\\
$^6$Astronomical Observatory, Jagiellonian University, ul. Orla 171,
30--244 Cracow, Poland, E-mail: siwak@oa.uj.edu.pl\\
$^7$Astrophysikalisches Institut und Universit\"ats-Sternwarte,
Schillerg\"asschen 2-3, D-07740 Jena, Germany\\
}

\date{Accepted 0000 Month 00. Received 0000 Month 00; in original form 2007 March 17}

\pagerange{\pageref{firstpage}--\pageref{lastpage}} \pubyear{2008}

\maketitle

\label{firstpage}

\begin{abstract}

Tightest known quadruple systems VW~LMi consists of contact eclipsing binary with 
$P_{12}$ = 0.477551 days and detached binary with $P_{34}$ = 7.93063 days 
revolving in rather tight, 355.0-days orbit. This paper presents new photometric 
and spectroscopic observations yielding 69 times of minima and 36 disentangled
radial velocities for the component stars. All available radial velocities and 
minima times are combined to better characterize the orbits and to derive absolute 
parameters of components. The total mass of the quadruple system was estimated at 
4.56 M$_\odot$. The detached, non-eclipsing binary with orbital period $P$ = 7.93 days 
is found to show apsidal motion with $U \approx 80$ years. Precession period in this 
binary, caused by the gravitational perturbation of the contact binary, is estimated 
to be about 120 years. The wide mutual orbit and orbit of the non-eclipsing pair are 
found to be close to coplanarity, preventing any changes of the inclination angle 
of the non-eclipsing orbit and excluding occurrence of the second system of eclipses 
in future. Possibilities of astrometric solution and direct resolving of the wide, 
mutual orbit are discussed. Nearby star, HD95606, was found to form loose binary 
with quadruple system VW~LMi. 
\end{abstract}


\section{INTRODUCTION}
\label{intro}

The photometric variability of VW~LMi (HIP~54003, HD~95660, sp. type F3-5V, 
$V_{max}$=8.0), was found by the Hipparcos mission \citep{hipp}, where it 
was correctly classified as a W~UMa-type eclipsing binary with an orbital 
period of 0.477547 days. The first ground-based photometric observations of the system 
obtained in 1999 and 2000 (taken in the $B$ and $V$ Johnson filters) were 
published by \citet{dumi2000}. Analysis of these light curves \citep{dumi2003} 
lead to the photometric mass ratio $q_{ph}$ = 0.395 and inclination $i$ = 72.4\degr and 
contact configuration for the system. Later \citet{gome2003} presented 
new $BV$ photometry and its preliminary analysis. Assuming convective envelopes 
for both components and the temperature of the primary as $T_1$ = 6700 K the 
authors estimated  the mass ratio as $q=0.4$, and inclination around 70\degr. 
Fourier analysis of its Hipparcos light curve (hereafter LC) presented by \citet{sela2004} 
yielded quite different parameters: $q = 0.25$, $i$ = 72.5\degr~and fill-out factor 
$f=0.4$. The discovery of the second (non-eclipsing) binary in VW~LMi by \citet{ddo11} makes 
all previous photometric solutions almost useless due to strong light contribution of the
second pair of about $(L_3 + L_4)/(L_1 + L_2) = 0.42$ (at the maximum brightness 
of the contact pair) which was not taken into account. 

\citet{ddo11} presented long-term spectroscopy (209 spectra
taken between 1998 and 2005) of the system obtained at David Dunlap Observatory
(hereafter DDO) which enabled to disentangle all three orbits in this tight 
multiple system: the contact binary with the $P_{12}$ = 0.4775 days period 
is orbiting another binary with $P_{34} = 7.93$ days in a relatively tight, 355-days,
mutual orbit. Using preliminary inclination angle of 
the contact-binary orbit found by photometric analysis, $i_{12}$ = 80.1$\degr$, 
the authors determined masses of all components and found that the orbits of the 
binaries are not coplanar. Light-time effect (hereafter LITE) with peak-to-peak range 
of $2A$ = 0.0074 days was predicted to be seen in the minima of the contact pair 
as a result of the mutual revolution of the binaries. The LITE was found in 
published minima by \citet{bajo2007}. The corresponding orbital parameters, 
$A$ = 0.0037(4) days, $P_{1234}$ = 353(2) days, $e$ = 0.5(2) and 
$\omega$ = 2.8(4) rad, are rather preliminary due to few
available minima. The eccentricity corresponding to their orbital solution 
is much higher than predicted spectroscopically by \citet{ddo11}.

The spectral type of VW~LMi was estimated as F5V by \citet{ddo11}, while observed Tycho-2
$(B-V)$ = 0.21 and 2MASS $J-K$ = 0.34 colors correspond to F2V spectral type.
Both determined spectral type and colors refer to the whole quadruple system.

VW~LMi is tightest quadruple system known \citep{toko2008}. Also it has the 
shortest period of the outer orbit within multiple systems harboring 
contact binaries. The ratio of the outer orbital period and orbital period of
the non-eclipsing pair is only about $P_{1234}/P_{34}$ = 44.5 hence we can
expect secular orbital changes on the timescales as short as decades. 
The chances of resolving of components (binaries) by either speckle or long-baseline 
interferometry are rather meager: the maximum angular separation of the components
was estimated to be only about 10mas \citep{ddo11}. Astrometric observations of
VW~LMi do not indicate its multiplicity. 

Goals of the present paper are as follows: (i) to present and analyze new
photometric and spectroscopic observations, (ii) to perform simultaneous solution
of LITE using both minima times of the contact binary and
radial velocities (hereafter RVs) of the individual components, (iii) to assess 
possibility of the tidal disturbances of the inner orbits and resulting precession,
(iv) determine absolute parameters of all four components.

\section{New observations}
\subsection{Photoelectric and CCD photometry}
\label{photometry}

To secure as many minima times as possible, photometry was performed at several
observatories in Slovakia and Germany. Since the system was too bright for
most of the instruments equipped with CCD cameras, photolenses were extensively 
used and on larger telescopes, the observations were mostly performed in the 
Johnson $B$ filter which enabled reasonable exposure times. The information on 
individual observatories and instruments used is given in Table~\ref{tab01}; 
parameters of variable and comparison stars are comprehensively listed in 
Table~\ref{tab02}. 

\begin{table}
\caption{Overview of telescopes and instruments/detectors used to obtain 
         photometry of VW~LMi. Abbreviations of the observatories (Obs.):
         G1, G2 pavilions of the Star\'a Lesn\'a Observatory, GSH - 
         Grossschwabhausen observing station of the Jena University,
         KO - Astronomical Observatory at Kolonica saddle 
         (belongs to the Vihorlat Observatory), RO - Roztoky Observatory. 
         Instruments using a photolense at KO and RO are named using
         its focal length. \label{tab01}} 
\footnotesize
\begin{center}
\begin{tabular}{lclc}
\hline
\hline
Obs.    & Telescope &  Detector   &  Filters    \\
\hline
G1      & 508/2500  &  SBIG ST10-MXE & $UBV(RI)_C$ \\
G2      & 600/7500  &  EMI 9789QB    & $UBVN$      \\
GSH     & 250/2250  &  CTK102        & $B$         \\
KO180   & 60/180    &  MEADE DSI Pro & $N$         \\
KO400   & 80/400    &  MEADE DSI Pro & $VN$        \\
KO      & 300/2400  &  SBIG ST9-XE   & $B$         \\
RO200   & 100/200   &  MEADE DSI Pro & $N$         \\
\noalign{\smallskip}
\hline 
\hline
\end{tabular}
\end{center}
\end{table}

Photoelectric photometry of VW~LMi was obtained using
60cm Cassegrain telescope in the G2 pavilion of the Astronomical Institute of the Slovak
Academy of Sciences. VW~LMi was observed in the Johnson $BV$ filters and with
neutral filter (optical glass BK7, denoted in tables as $N$). $BV$ observations 
were transformed to the international photometric system. The magnitude differences
with respect to the comparison star HD95606 were corrected for the differential
extinction using nightly or seasonal extinction coefficients. Since the angular
distance of the comparison and variable star is only 5.65 arcmin, the differential
extinction correction was rather negligible, and never exceeded 0.002 mag. 
Occasionally observed check star, HD95527, showed good stability of the
comparison star, HD95606. 

\begin{table}
\caption{Parameters of the variable and comparison stars used. The
         proper motions and $(B-V) = 0.85(B-V)_T$ colors were taken from the Tycho-2 
         Catalog \citep{Tycho2}, parallaxes from the Hipparcos Catalogue 
         \citep{hipp}, infrared colors, $(J-K)$, from the 2MASS \citep{2MASS}. 
         Radial (RV) and rotational velocities ($v \sin i$), and spectral 
         types were determined from the DDO spectroscopy; for VW~LMi 
         the listed RV corresponds to the systemic velocity of the 
         whole quadruple. \label{tab02}}
\footnotesize
\begin{center}
\begin{tabular}{lcccc}
\hline
\hline
                     & VW~LMi      & cmp1        & cmp3        \\
\hline
GSC                  & 2519-2347   & 2519-1195   & 2519-1326   \\
HD                   & 95660       & 95606       & 95527       \\
HIP                  & 54003       & 53969       & --          \\
$\mu_\alpha \cos \delta$ [mas.y$^{-1}$]  & 12.7(11) & 10.7(11) & $-$9.8(12)  \\
$\mu_\delta$ [mas.y$^{-1}$]              & $-$5.0(12) & $-$3.4(11) &$-$67.9(12)  \\
$RV$ [km.s$^{-1}$]   & $-0.15$     & 7.80        & 11.10       \\
$v \sin i$ [km.s$^{-1}$] &  --     & 32          & $<13$       \\
$\pi$ [mas]          & 8.04(0.90)  & 7.07(1.25)  & --          \\
$(B-V)$              & 0.340(21)   & 0.398(28)   & 0.570(34)   \\
$(J-K)$              & 0.208(30)   & 0.266(31)   & 0.300(35)   \\
\hline
sp. type             &  F3-5V      & F5V         & G0V         \\
\noalign{\smallskip}
\hline
\hline
\end{tabular}
\end{center}
\end{table}

Most of the CCD photometry was secured using 50cm Newtonian telescope in the G1 pavilion 
of the Astronomical Institute of the Slovak Academy of Sciences. Practically all observations 
were taken through the Johnson $B$ filter, only on March 25, 2007 
$UBV(RI)_C$ photometry was performed. The data were left in the instrumental system 
since the major purpose was securing minima instants. Further large set of data was 
obtained using various small telescopes/photolenses at the Astronomical Observatory at Kolonica saddle
(48\degr 56' 6" N, 22\degr 16' 26"E). Several times of minima were also obtained at 
the Roztoky Observatory (49\degr 33' 57" N, 21\degr 28' 54"E). CCD photometry of VW~LMi 
was also obtained at Grossschwabhausen observing station of the Jena University Observatory
(see \citet{mugr2008}). CCD reduction was carried out by usual way (bias, dark, 
flat-field corrections) and aperture photometry was performed by procedures in MIDAS 
environment (only at Star\'a Lesn\'a Observatory), and C-Munipack 
package\footnote{http://munipack.astronomy.cz/}. HD95606 served as comparison star 
at all observatories. 

\subsection{Medium dispersion spectroscopy}
\label{spectroscopy}

New spectroscopic observations were obtained using the slit spectrograph in 
the Cassegrain focus of 1.88m telescope of DDO. The observations were performed 
between March 2007 and very beginning of July 2008, when DDO ceased to operate. 
The spectra were taken in a window of about 240~\AA~ around the Mg~I triplet 
(5167, 5173 and 5184~\AA) with an effective resolving power of 
$R$ = 12,000 - 14,000. The diffraction grating with 2160 lines/mm was used 
with the sampling of 0.117 A/pixel. One-dimensional spectra were extracted by 
the usual procedures within the IRAF environment\footnote{IRAF is distributed 
by the National Optical Astronomy Observatories, which are operated by the 
Association of Universities for Research in Astronomy, Inc., under cooperative 
agreement with the NSF.} after the bias subtraction and the flat-field division. 
Cosmic-ray trails were removed using a program of \citet{pych2004}. 

\begin{figure}
\includegraphics[width=80mm,clip=]{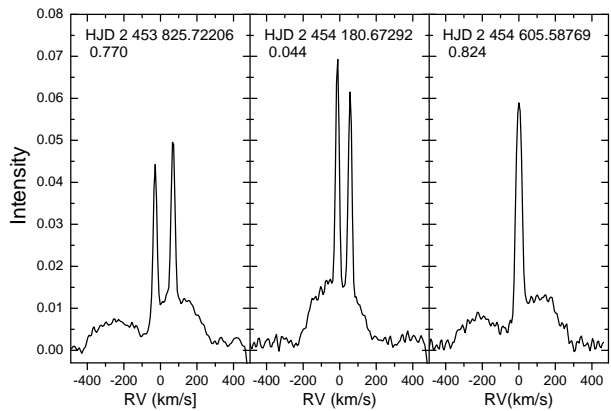}
\caption{Broadening functions of VW~LMi. From left to right: all components
visible, contact binary close to the conjunction, and detached binary close to the
conjunction. Original broadening functions were smoothed convolving with
a Gaussian function ($\sigma$ = 10 km~s$^{-1}$). Heliocentric Julian dates 
and phases of the contact binary are given in individual panels. \label{fig01}}
\end{figure}

All exposures were 900 seconds long and lead to S/N of about 100-150.
The observations were analyzed using the technique of broadening-functions, 
described in \citet{Rci1992,Rci2002}. For VW~LMi we used HD222368 
as the template (F7V, $v \sin i = 3$ km~s$^{-1}$). This star provided 
the best match to the observed spectra - integral of BF was always 
0.95 - 1.00\footnote{Late spectral type of the best template indicates 
later spectral type of the second binary in VW~LMi}. Similar to 
\citet{ddo11} we firstly fitted simple multi-component Gaussian model to 
the extracted BFs (see Fig.~\ref{fig01}), then detached pair with slowly-rotating components was subtracted. 
The wide features corresponding to the contact pair were subsequently modelled by 
the rotational profiles which effectively corresponds to approximation of the 
components of the contact binary by rotating limb-darkened spheres.
Since shapes of the profiles slightly depend on the limb darkening we fixed
$u = 0.53$ according to the tables of \citet{hamme93} appropriate for main-sequence
F3V star at 5184 \AA. RVs determined from the new spectra are 
given in Table~\ref{tab03}. Several BFs extracted from observations taken close
to conjunction of the contact binary did not allow to
determine their RVs. In case of non-eclipsing binary, RVs of its components
could be reliably determined from all new spectra.

\begin{table}
\caption{New radial velocities for all four components of the quadruple system
         VW~LMi. While radial velocities of the detached pair (RV3,RV4; $P$ = 7.93 days)
         were determined by multi-component Gaussian fitting to the observed
         BFs, radial velocities of the contact binary (RV1,RV2; $P$ = 0.4775 days)
         were determined by rotational-profile fitting to the residual BFs
         after removal of the detached pair from BFs.  Radial velocities
         RV1, RV2 are not given for observations performed close to the conjunctions.
         Heliocentric Julian dates were not corrected for the LITE
         on the 355-days orbit. Components denoted as ``1'' and ``3'' are more massive 
         ones in either of the binaries.
         \label{tab03}}
\footnotesize
\begin{center}
\begin{tabular}{lrrrr}
\hline
\hline
 HJD         &   RV1    &   RV2      &    RV3    &   RV4    \\
2\,400\,000+ & [km/s]   & [km/s]     &  [km/s]   & [km/s]   \\
\hline
 53823.67103 &    --    &    --      & $-16.88$  &   62.46  \\
 53825.72206 &   84.55  & $-245.57$  &   68.77   & $-29.03$ \\
 53825.73278 &   78.31  & $-237.41$  &   69.77   & $-28.65$ \\
 53825.74469 &   78.44  & $-230.85$  &   69.92   & $-29.05$ \\
 53825.75545 &   72.69  & $-206.33$  &   70.55   & $-28.90$ \\
 53835.80611 &    --    &    --      &   53.04   & $-15.75$ \\
 53836.77953 &    --    &    --      &    3.88   &   36.12  \\
 53836.78848 &    --    &    --      &    3.86   &   36.66  \\
 54167.69771 &   68.79  & $-188.87$  &   87.41   & $-39.54$ \\
 54167.70691 &   58.05  & $-180.22$  &   87.23   & $-39.55$ \\
 54177.69983 &   71.34  & $-231.89$  &   12.57   &   34.40  \\
 54179.77080 &$-112.25$ &   173.30   & $-40.74$  &   84.99  \\
 54180.67292 &    --    &    --      & $-11.94$  &   57.31  \\
 54182.66132 &$-114.17$ &   203.01   & $-27.91$  &   73.79  \\
 54183.65132 &$-118.18$ &   213.28   & $-42.67$  &   83.14  \\
 54189.64139 &   72.54  & $-210.75$  &   34.34   &    6.43  \\
 54190.66416 &    --    &    --      & $-33.23$  &   71.64  \\
 54193.67997 &$-108.38$ &   231.24   &   37.19   &    3.20  \\
 54193.75446 &    --    &    --      &  $-0.17$  &   41.89  \\
 54200.62450 &   81.57  & $-220.56$  &   49.60   & $-17.90$ \\
 54507.71485 &   58.29  & $-204.99$  &   69.31   & $-26.19$ \\
 54515.82080 &   64.89  & $-205.72$  &   75.41   & $-31.01$ \\
 54518.69236 &   61.91  & $-195.11$  &   13.48   &   31.94  \\
 54521.78963 & $-99.01$ &   170.49   &  $-4.12$  &   56.24  \\
 54542.78635 &$-104.76$ &   193.21   &  $-6.60$  &   44.71  \\
 54546.74822 &   70.95  & $-166.15$  &   43.40   &  $-5.20$ \\
 54550.69995 &   67.16  & $-184.81$  &  $-4.91$  &   43.48  \\
 54605.58769 &  101.56  & $-217.66$  &    7.86   & $-10.86$ \\
 54619.58209 & $-67.07$ &   187.80   &   56.99   & $-68.57$ \\
 54626.64631 &    --    &    --      &   38.25   & $-55.24$ \\
 54626.65701 &    --    &    --      &   38.87   & $-55.73$ \\
 54628.60047 &    --    &    --      &   34.47   & $-55.86$ \\
 54638.59131 &    --    &    --      & $-65.05$  &   40.82  \\
 54642.60165 & $-79.75$ &   234.95   &   37.10   & $-62.29$ \\
 54647.59897 &  121.47  & $-208.31$  & $-77.27$  &   51.63  \\ 
 54649.58784 &    --    &    --      &  $-2.21$  & $-28.02$ \\
\noalign{\smallskip}
\hline
\hline
\end{tabular}
\end{center}
\end{table}

\section{Minima determination}
\label{minima}

Most widespread approach to obtain instants of minima of eclipsing binaries is
to use Kwee \& van Woerden's method \citep{kw1956}. From our experience the errors
estimated using their formula (14) are often unrealistically small. The real
uncertainties are very probably dominated by systematic errors. The principal
problem in case of CCD photometry is scattered light which cannot be fully corrected 
by flatfielding. This results in spurious shifts/trends in differential photometry
in case that the field is not perfectly guided making times of minima systematically
shifted\footnote{Unfortunately, neither of the telescopes uses an autoguider for
pointing}. Our photometry could possibly be improved by using an algorithm 
based on Principal Component Analysis proposed by \citet{tamuz2005}. Unfortunately,
the frames were obtained at several observatories with different setups and even
different orientation of the field. Systematic errors in minima positions
were partially removed by fitting technique proposed below.

\begin{table*}
\caption{New times of minima of VW~LMi determined by the template fitting to
         observed LCs. Observations in individual filters were treated
         separately. If both minima were observed in given night (indicated in 
         column ``Type'') the instant given in the table refers to the first one.
         The standard errors of the last digit are given in parentheses.
         \label{tab04}}
\footnotesize
\begin{center}
\begin{tabular}{lccl|lccl|lccl}
\hline
\hline
HJD             & Fil.& Obs.&Type& HJD             & Fil.& Obs.&Type& HJD             & Fil.& Obs.& Type \\
2\,400\,000+    &     &     &    & 2\,400\,000+    &     &     &    & 2\,400\,000+    &     &     &      \\
\hline
53461.45927(21) & $B$ & G2  & I  & 54063.65646(14) & $B$ & G1    & I  & 54245.35634(11) & $B$ & G1     & II  \\
53461.45883(16) & $V$ & G2  & I  & 54066.52319(13) & $B$ & G1    & I  & 54415.61257(30) & $V$ & KO     & I   \\
53461.45969(20) & $N$ & G2  & I  & 54085.62425(21) & $V$ & KO400 & I  & 54469.57672(29) & $B$ & G1     & I   \\
53465.51832(22) & $B$ & G2  & II & 54088.49087(24) & $B$ & G1    & I  & 54469.57628(29) & $V$ & G1     & I   \\
53465.51792(15) & $V$ & G2  & II & 54095.65269(10) & $B$ & G1    & I  & 54469.57549(20) & $N$ & KO180  & I   \\
53465.51843(22) & $N$ & G2  & II & 54114.51541(08) & $B$ & G1    & II & 54499.66040(07) & $B$ & GSH    & I   \\
53767.57388(15) & $N$ & G2  & I  & 54116.66409(07) & $B$ & G1    & I  & 54505.38874(26) & $N$ & KO400  & I   \\
53791.44941(26) & $B$ & G2  & I  & 54117.61952(14) & $N$ & KO400 & I  & 54509.44938(43) & $N$ & KO400  & II  \\
53791.44924(21) & $V$ & G2  & I  & 54148.65937(14) & $B$ & G1    & I  & 54512.55203(10) & $B$ & GSH    & I   \\
53791.45004(33) & $N$ & G2  & I  & 54148.65953(35) & $V$ & KO400 & I  & 54521.62540(08) & $V$ & GSH    & I   \\
53794.31433(08) & $B$ & G1  & I  & 54149.61531(13) & $B$ & G1    & I  & 54532.36983(15) & $N$ & KO400  & II,I\\
53797.41955(16) & $B$ & G1  & II & 54162.50371(27) & $V$ & KO400 & I  & 54556.48551(26) & $V$ & KO400  & I   \\
53802.43347(31) & $B$ & G2  & I  & 54167.28101(36) & $V$ & KO400 & I  & 54557.43985(19) & $N$ & RO     & I   \\
53802.43276(22) & $V$ & G2  & I  & 54172.53421(10) & $B$ & G1    & I  & 54571.28992(25) & $N$ & RO     & I,II\\
53802.43260(24) & $N$ & G2  & I  & 54173.48896(12) & $B$ & G1    & I  & 54581.31655(14) & $N$ & RO     & I,II\\
53814.37177(07) & $B$ & G1  & I  & 54173.48891(17) & $V$ & KO400 & I  & 54582.51234(24) & $N$ & RO     & II  \\
53815.32588(05) & $B$ & G1  & I  & 54175.39958(16) & $V$ & KO400 & I  & 54584.41972(65) & $N$ & KO400  & II  \\
53815.32524(08) & $N$ & G2  & I  & 54185.42614(26) & $B$ & G1    & I  & 54584.42043(18) & $N$ & RO     & II  \\
53833.47072(07) & $B$ & G1  & I  & 54185.42681(32) & $V$ & G1    & I  & 54585.37523(15) & $N$ & RO     & II  \\
53845.40960(09) & $B$ & G1  & I  & 54185.42691(41) & $R$ & G1    & I  & 54586.32887(24) & $B$ & GSH    & II  \\
53850.42278(11) & $B$ & G1  & II & 54185.42709(39) & $I$ & G1    & I  & 54588.47856(17) & $B$ & GSH    & I   \\
53867.37715(08) & $B$ & G1  & I  & 54185.42676(24) & $V$ & KO400 & I  & 54593.49393(18) & $N$ & RO     & II  \\
53894.35622(16) & $B$ & G1  & II & 54189.48500(30) & $V$ & KO400 & II & 54594.44843(18) & $N$ & RO     & II  \\
54026.64630(10) & $B$ & G1  & II & 54191.39733(21) & $V$ & KO400 & II & 54595.40379(19) & $V$ & KO400  & II  \\
54027.60191(19) & $B$ & G1  & II & 54195.45500(15) & $V$ & KO400 & I  & 54599.46132(19) & $B$ & GSH    & I   \\
54057.68914(14) & $B$ & G1  & II & 54212.40823(10) & $B$ & G1    & II & 54615.46100(25) & $N$ & RO     & II  \\
54059.59910(10) & $B$ & G1  & II & 54213.36179(07) & $B$ & G1    & II & 54616.41716(20) & $N$ & RO     & II  \\
\hline
\hline
\end{tabular}
\end{center}
\end{table*}

Since LC of the system appears to be very stable (we do not see any asymmetries 
or changes) fitting templates were prepared to obtain instant of conjunction (minimum)
for any sufficiently long photometric sequence. Such a way we made use not 
only of the minima but of other LC segments. The template LCs were represented 
by symmetric trigonometric series of the 10th order. Even if the eclipsing pair is 
a contact binary, amplitude of its LC depends on the wavelength. Using CCD 
photometry of March 25, 2007 we see the following amplitudes $\Delta B$ = 0.46, 
$\Delta V$ = 0.43, $\Delta R_c$ = 0.41, and $\Delta I_c$ = 0.38. The differences in 
amplitude primarily result from wavelength-dependent limb darkening and light 
contribution of the third component (see Section \ref{lc}). Due to the differences 
in filter transparencies and wavelength response of detectors we had to form a 
template LC for each filter separately and the fitting LC was scaled to match the
observations. We also noted small ($\approx$ 0.02 mag) shifts of the LCs observed 
even with the same instrument. Sometimes even slight slopes of the LC were recorded. 
These shifts/slopes are very probably caused by scattered light combined with different 
pointing of the telescope. 

To obtain good fits of the observed LCs by templates $T(x)$ we formed the 
following fitting function:

\begin{equation}
F(x) = A + B.x + C.T(x-D).
\end{equation}

This allowed for shifting, scaling and ``slanting'' of the template LC. Fixing of the 
parameters was judged according to the appearance of individual LCs. All new times of 
minima are listed in Table \ref{tab04}. The formal errors given are still much
smaller than systematic errors. Times of ``minima'' determined from sections of LC
not containing minima were omitted from further analysis because of significantly
higher errors.

\begin{figure}
\includegraphics[width=80mm,clip=]{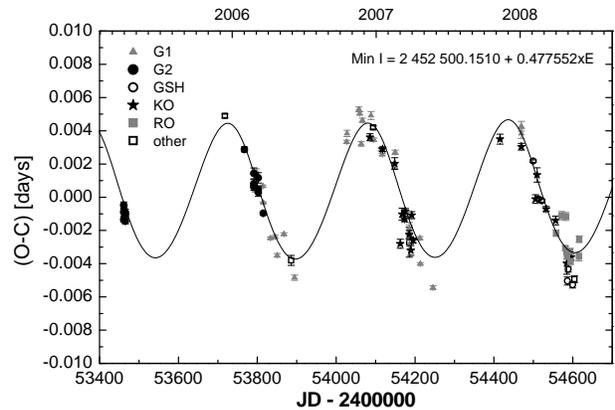}
\caption{(O-C) diagram of VW~LMi showing best covered interval since 2005.
 The (O-C) data were given with respect to the following ephemeris for the 
 primary minimum: 2\,452\,500.1510 + 0.477552$\times$E. The primary and secondary
 minima were plotted using the same symbols. The optimum fit corresponding
 to Table~\ref{tab05} is plotted. \label{fig02}}
\end{figure}

The resulting (O-C) diagram including published minima times is shown in 
Fig.~\ref{fig02}. The minima times of VW~LMi can be found in 
\citet{gome2003,dumi2003,derm2003,por2005,dwor2005,dwor2006,hub2007,nel2007}
and there are four unpublished CCD times of minima observed by 
K. Nakajima\footnote{see http://www.kusastro.kyoto-u.ac.jp/vsnet/} - 
HJD 2\,453\,461.9371(2), 2\,453\,462.1758(2), 2\,453\,886.0001(3), and 2\,454\,603.9984(2).
The residuals, from rather arbitrary ephemeris given in the figure, clearly 
show wave-like behavior with one-year periodicity.

\section{Global fit of the data}
\label{rvlite}

Multiple-dataset fitting is rather widespread technique nowadays. It is typical
to combine RVs and visual orbits or RVs and interferometric visibilities of binary stars
(e.g., \citet{pour1999}, \citet{tang2006}). Eclipsing-binary modelling programs
(see \citet{hamm2007}) enable to combine RV data LCs to determine third-body
parameters. The case of VW~LMi is, however, special and unprecedented, requiring 
ad-hoc approach.

Since our photometry was performed at several observatories, often without
filter, we decided not to fit a global model to LCs together with RVs and to
use only minima times to better characterize the outer orbit. Rigorous modelling 
of those datasets (four RV curves and times of minima) is
quite complex since times of all RV observations should be 
properly corrected for the LITE caused by the mutual revolution of the binaries.
This effect is minor in the non-eclipsing pair, where amplitude of LITE 
is about 0.004 days, which amounts to only about $\pm$0.0005 in phase. In case of eclipsing 
pair the amplitude of LITE is about 0.0037 days, and $P_{12}$ = 0.4775 days make 
for $\pm$0.0077 phase shifts which cannot  fully be neglected. In fact, correcting 
observed times of RV measurements for the LITE significantly decreased global
$\chi^2$. LITE is also manifested directly in observed times of minima. The 
informational contents of the minima is, unfortunately, very similar to having 
observed only RVs of the eclipsing pair. Integrals of RV curves are radial distances from the observer corresponding directly to the LITE 
through the finite speed of light. Both datasets are, however, nicely 
complementary: during the spectroscopic conjunction, when the RVs 
of the components are not measurable, we can observe minima giving the radial 
distance of the eclipsing binary in the outer orbit. Moreover, the precise minima 
times are easy to observe using a small telescope or photolense equipped with a 
cheap CCD camera. As we will see later, the relative orbit of mass center of 
the contact binary is much better defined by the LITE than RVs. This is crucial 
to establish mass ratio of binaries, inclinations and masses of components.

In further description we will use similar notation of parameters as in \citet{ddo11}:
common parameters (orbital period, eccentricity, time of periastron passage, 
longitude of periastron) for orbits will be denoted by index ``12'' (contact 
binary), ``34'' (detached non-eclipsing binary) and ``1234'' (mutual orbit).
Semi-amplitudes of RV changes for the individual stars will be denoted
by separate index ``1'', ``2'', ``3'', or ``4'' while semi-amplitudes of systemic
velocity changes of both binaries as whole in the mutual orbit will be 
denoted as ``12'' and ``34''. Finally, systemic velocity of the mass center of the whole
quadruple as $V_0$. For RVs of components of contact binary we 
can write:

\begin{eqnarray}
RV_i = V_0 + K_{12} [e_{1234} \cos \omega_{1234} + \cos (\nu_{1234} + \omega_{1234})]+ \nonumber \\
       +(-1)^{i+1} K_i [e_{12} \cos \omega_{12} + \cos (\nu_{12} + \omega_{12})]      
\end{eqnarray}

where index $i =1,2$ denotes the component. For the detached binary we similarly
have:

\begin{eqnarray}
RV_i = V_0 - K_{34} [e_{1234} \cos \omega_{1234} + \cos (\nu_{1234} + \omega_{1234})] + \nonumber \\
       +(-1)^{i+1} K_i [e_{34} \cos \omega_{34} + \cos (\nu_{34} + \omega_{34})].
\end{eqnarray}

with $i=3,4$. True anomalies of inner orbits ($\nu_{12}, \nu_{34}$) have to be 
corrected for the LITE caused by the mutual revolution of binaries. 
If we accept the plane parallel to the sky and intersecting mass center of the
quadruple system as the reference plane, then LITE seen in either 
of the binaries is:

\begin{equation}
\Delta T_j = \frac{K_j P_{1234} (1-e_{1234})^{3/2}}{2\pi c} \frac{\sin(\nu_{1234}+
           \omega_{1234})}{1 + e_{1234} \cos \nu_{1234}},
\end{equation} 

where index $j$ is either ``12'' or ``34''. In case of the eclipsing pair, where the
orbital eccentricity is zero, the predicted times of minima occurring at epoch 
$E$ would be:

\begin{equation}
T_{min} = T_{12} + P_{12} E + Q E^2 + \Delta T_{12} 
\end{equation}

if true anomalies in the eclipsing binary orbit are counted from the upper
conjunction of the more massive component (then $\omega_{12} = \pi/2$ with
$e_{12}$ = 0). 

While amplitude of the RV changes of the systemic velocity
of the detached-pair orbit around the common center of gravity is rather
reliable due to very well defined sharp peaks in BFs ($v \sin i < 12$ km~s$^{-1}$ 
which is limit given by the spectral resolution), the RVs of contact-binary components
are rather imprecise due to a ``cross-talk'' with the second binary and wide
profiles. This makes size of the relative orbit of the contact binary 
(related to $K_{12}$) rather unreliable. The shape of the orbit is, fortunately,
coupled through well defined orbit of the detached pair. The size of the 
contact-binary orbit is much-better defined by LITE seen in the observed minima.

The best fits to the data (4 RV curves and the LITE orbit) were performed
using differential corrections method using analytic derivatives of the functions.
The dependence of true anomaly on mean anomaly ($M$) and eccentricity 
was represented in derivatives by the truncated series up to the
second degree in orbital eccentricity:

\begin{equation}
\nu \approx M + 2e \sin M + \frac{5}{4} e^2 \sin 2M.
\end{equation}

Since all orbits are being close to circular this approximation was sufficient to
make the optimization process quickly convergent. The sum of {\it normalized} $\chi^2$ was
used as the merit function:

\begin{equation}
S = \sum_{j=1}^{5} \frac {\sum_{i=1}^{n_j} (O_{ij} - C_{ij})^2}{n_j \sigma_j^2}
= \sum_{j=1}^{5} \chi^2_j
\end{equation}

where $j$ is index of the dataset, $n_j$ is number of observation for the given
dataset, $i$ is index of individual observation and $\sigma_j$ is average standard 
deviation of individual datapoint for the $j$-th dataset. Average uncertainty of
a datapoint for individual datasets was estimated fitting separately times
of minima and RVs. For minima times we got $\sigma_{min}$ = 0.0008 day (which is significantly
more than typical error in Table~\ref{tab04}), for 
RVs $\sigma_{RV1}$ = 11.3 km~s$^{-1}$, $\sigma_{RV2}$ = 13.1 km~s$^{-1}$,
$\sigma_{RV3}$ = 2.6 km~s$^{-1}$, and $\sigma_{RV4}$ = 2.2 km~s$^{-1}$.
The estimated average uncertainties were found to be realistic since 
normalized $\chi^2$ for individual datasets were close to unity. Observations within 
individual datasets were assigned the same errors (as given above) due to the 
fact that both times of minima and RVs are mostly affected by systematic errors 
which cannot  reliably be determined. One minimum (2\,450,877.4658), published 
by \citet{gome2003}, giving very large (O-C) (probably due to a typing error), 
was omitted from further analysis. 

\begin{figure}
\includegraphics[width=80mm,clip=]{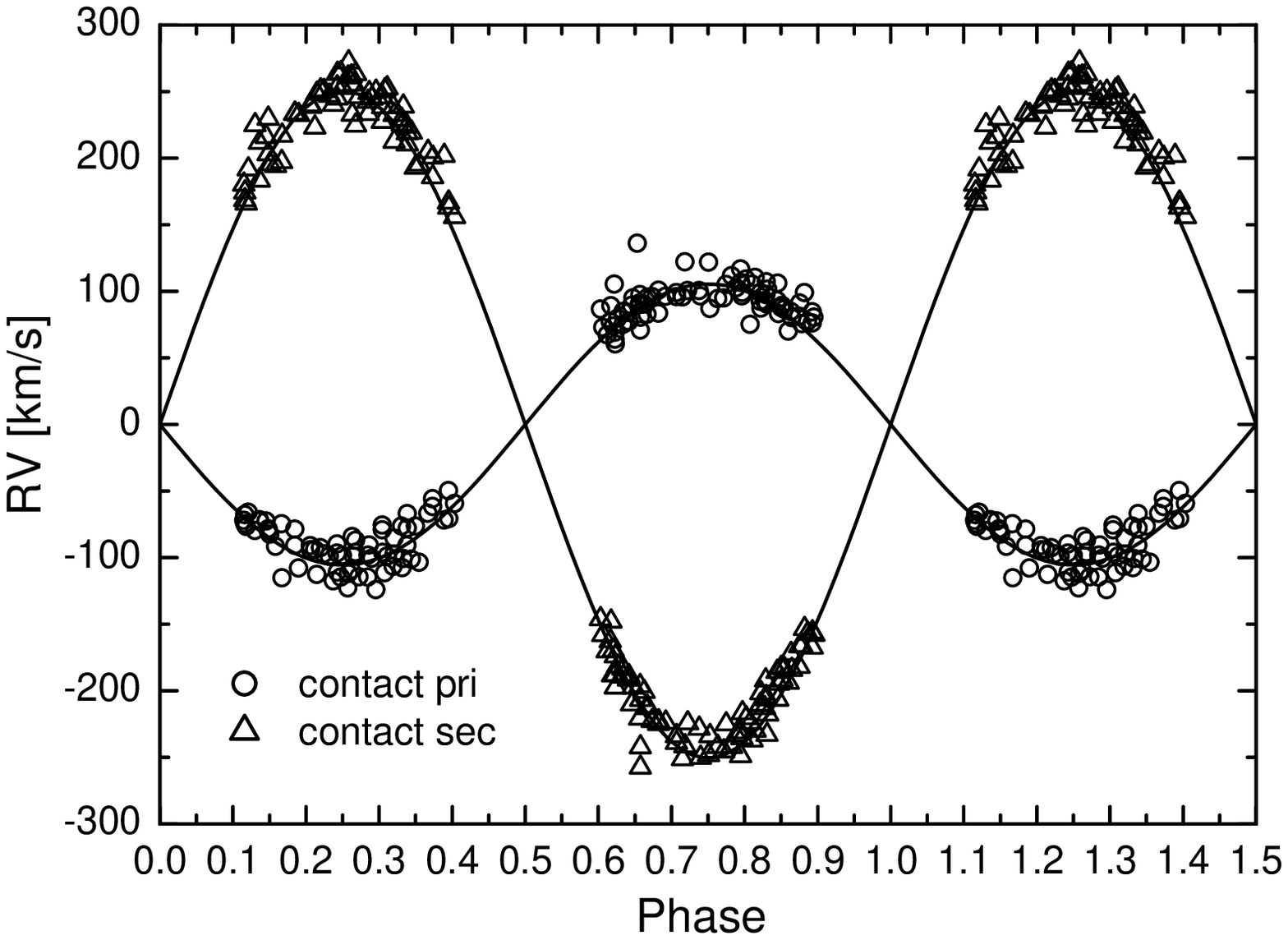}
\includegraphics[width=80mm,clip=]{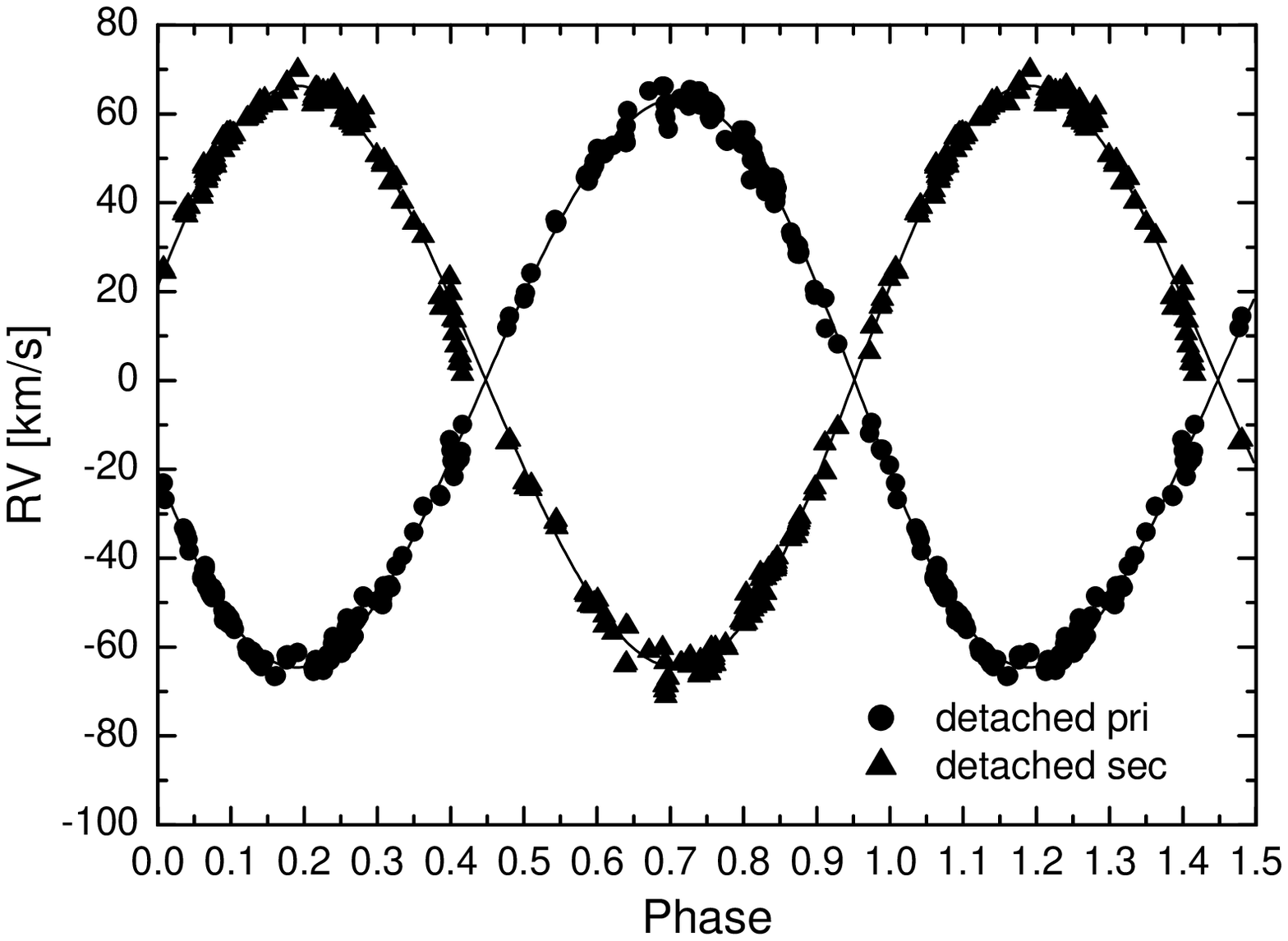}
\includegraphics[width=80mm,clip=]{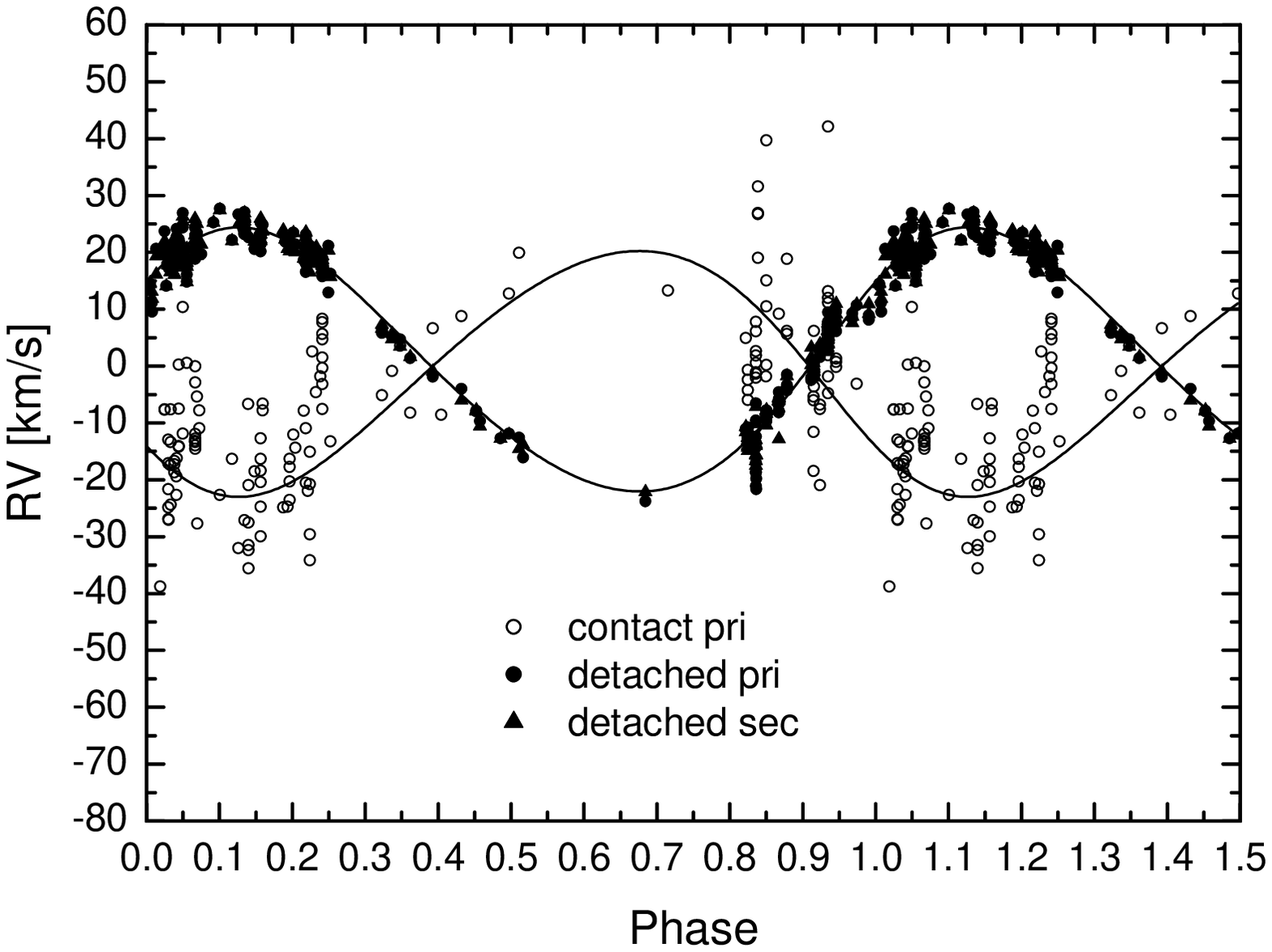}
\caption{Radial velocities of all four components and their best fits 
         assuming hierarchic quadruple model. Phases of the measurements
         in inner orbits were corrected for LITE in the outer
         orbit. Longitude of the periastron of the detached pair is assumed 
         to be constant. Top panel and center panels show RVs of 
         the contact and detached pairs, respectively, after correction
         for the revolution in the outer 355-days orbit. The bottom
         panel displays velocities of individual components corrected for motion
         in inner orbits - reflecting mass-center revolution (observations
         of the secondary component of the contact binary were not plotted for
         clarity). Because the 
         outer orbit period is close to one year, the seasonal observing 
         interval only slowly moved in the orbital phases between 1998 
         and 2008 and all phases have not been covered yet (bottom). 
         The best fits corresponding to the global solution (Table~\ref{tab05})
         are plotted. \label{fig03}}
\end{figure}

Fitting theoretical curves to five datasets required 18 parameters. 
It is interesting to note, that LITE of the eclipsing pair required only one
additional parameter - $Q$, due to the fact that the orbital period of the
contact binary was found to be continuously increasing. Since orbital eccentricity of the 
detached binary was found to be only $e_{34}$ = 0.03, we also considered a circular 
orbit. This, however, lead to substantially worse fit to the data. 
Resulting parameters are given in Table~\ref{tab05}. Corresponding 
fits are shown in Figs. \ref{fig02} and \ref{fig03}.

\begin{table}
\caption{Simultaneous fit to radial velocities of both binaries 
         and times of minima of the eclipsing pair. The designation of parameters 
         is as follows: the index ``12'' refers to the orbit of the contact pair, 
         while the index ``34'' refers to the orbit of the second, detached binary.
         Parameters of the mutual orbit of these binaries are indexed as ``1234''.
         The orbit of the contact pair is assumed to be circular 
         ($e_{12}$=0, $\omega_{12} = \pi/2$). Masses of the systems
         are abbreviated as $M_{1234} = M_1 + M_2 + M_3 + M_4, M_{12} = M_1 + M_2,
         M_{34} = M_3 + M_4$. Standard errors of parameters are given in parentheses 
         \label{tab05}}
\footnotesize
\begin{center}
\begin{tabular}{lcc}
\hline
\hline
\multicolumn{2}{l}{\bf Contact (eclipsing) pair - circular orbit} \\
$P_{12}$ [days]      & 0.47755106(3)      \\
$Q$ [days]           & 1.63(9)~10$^{-10}$ \\
$T_{12}$ [HJD]       & 2,452,500.1497(2)  \\
$K_1$ [km~s$^{-1}$]  &   105.8(1.0)       \\
$K_2$ [km~s$^{-1}$]  &   250.2(1.2)       \\
$M_{12} \sin^3 i_{12}$ [M$_\odot$]     &  2.231(23) \\
$\chi^2$(RV1)        &    1.086           \\
$\chi^2$(RV2)        &    1.058           \\
$\chi^2$(MIN)        &    1.086           \\
\hline
\multicolumn{2}{l}{\bf Detached (non-eclipsing) pair}\\
$P_{34}$ [days]      & 7.93063(3)         \\
$e_{34}$             & 0.035(3)           \\
$\omega_{34}$ [rad]  & 1.90(9)           \\
$T_{34}$ [HJD]       & 2,452,274.54(11)   \\
$K_3$ [km~s$^{-1}$]  & 63.99(23)          \\
$K_4$ [km~s$^{-1}$]  & 65.53(27)          \\
$M_{34} \sin^3 i_{34}$ [M$_\odot$]      &  1.785(11) \\
$\chi^2$(RV3)        &    0.862           \\
$\chi^2$(RV4)        &    0.851           \\
\hline
\multicolumn{2}{l}{\bf Mutual wide orbit} \\
$P_{1234}$ [days]      & 355.02(17)       \\
$e_{1234}$             &   0.097(11)      \\
$\omega_{1234} $ [rad] &   2.20(12)       \\
$T_{1234}$ [HJD]       & 2,453,046(6)     \\
$K_{12}$ [km~s$^{-1}$] & 21.61(49)        \\
$K_{34}$ [km~s$^{-1}$] & 23.22(33)        \\
$M_{1234} \sin^3 i_{1234}$ [M$_\odot$] &   3.32(10)  \\
\hline
$V_0$ [km~s$^{-1}$]    & $-$0.15(25)      \\
\hline
\hline
\end{tabular}
\end{center}
\end{table}

Unfortunately the orbital period of the mutual orbit is close to one year. The system can
be observed for minima from beginning of November till beginning of June. The
observing interval to get spectrum just before dawn and just after twilight is
about a month longer. Therefore, to cover the whole outer orbit, the system would have to
be monitored at least 12-15 years (for ground-based observations). Until then
parameters depending on overall shape of the RV curve (all except orbital period)
cannot be reliably determined. This affects reliable determination of mass ratio
of the binaries $M_{34}/M_{12} = K_{12}/K_{34}$ and projected total mass
$(M_1 + M_2 + M_3 + M_4) \sin^3 i_{1234}$.

\section{Light-curve solution}
\label{lc}

The eclipsing pair revolves in rather tight orbit around another binary which 
results in phase shifts of about $\pm$0.008 which cannot be neglected as seen in 
long-term modelling of RVs (see Section \ref{rvlite}). Therefore,
we used photometry from a short interval compared to mutual, 355-days orbit. 
To determine photometric elements of the eclipsing pair we used $BV$ photoelectric 
LCs obtained on March 31 and April 4, 2005 at the Star\'a Lesn\'a 
observatory giving full coverage of the phases in relatively short time compared 
to the mutual orbital period (about 1.2\%). The orbital period was fixed at
0.477551 days but time of upper conjunction of the more massive component 
was adjusted. The third light, $(L_3 + L_4)/(L_1 + L_2) = 0.42$, is known from 
spectroscopy. It was derived from a 240\AA~section of the spectrum centered at 5184\AA~
fairly close to the Johnson $V$ filter. General experience with BF technique is, however, 
that the third light is often overestimated \citep{ruci2008} due to the continuum
normalization problems of contact binaries and difference of spectral types of
additional components.

\begin{figure}
\includegraphics[width=80mm,clip=]{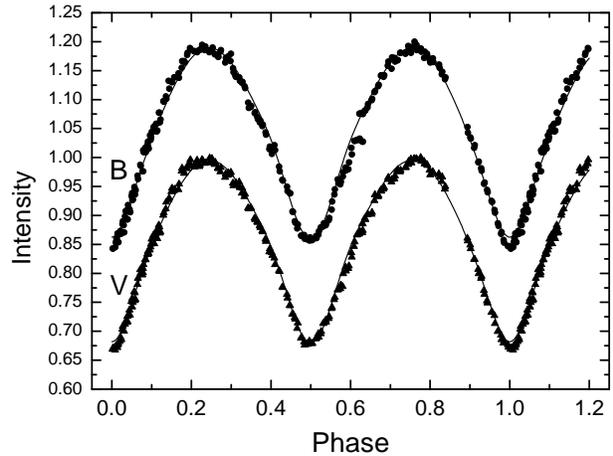}
\caption{Best fits to the $BV$ LCs for the adopted solution with 
         $i = 79.0$\degr~fixed. \label{fig04}}
\end{figure}

Analysis of LCs was performed using code {\it ROCHE} described in \citet{prib2004}.
Since eclipsing pair is a contact binary, circular orbit and synchronous rotation 
($F_1 = F_2 = 1$) was assumed in all fits. The model-atmosphere grids 
(see \cite{leje1997}) were selected for the solar metallicity. The temperature 
of the primary component was fixed according to its spectral type, F5V \citep{ddo11}, 
at 6460 K using calibration of \citet{popp1980}. Convective transfer of energy 
in the common envelope was assumed ($A_1=A_2=0.5, g_1=g_2=0.32$), and limb-darkening was
automatically interpolated from the extensive tables of \citet{hamme93} for mean
$\log g$ and $T_{eff}$ of the components.  Due to the fact that exact values of 
third light for $B$ and $V$ passbands are unknown, unlike in usual approach, we
performed several solutions with the orbital inclination fixed between 75 to 
85\degr. The mass ratio was not adjusted since it tightly correlates with
orbital inclination, and third light, but was rather fixed at spectroscopically determined
value $q_{sp}$ = 0.423. Optimization of the parameters lead, as expected, to contact 
configuration for every trial inclination. Surprisingly, third light was found to be 
almost identical in $B$ and $V$ passbands, indicating very similar spectral type of the
second non-eclipsing pair. For $i = 79.0 \degr$, third light in both passbands
was close to spectroscopically determined value $L_{34}/L_{12}(B)$ = 0.41$\pm$0.03
and $L_{34}/L_{12}(V)$ = 0.42$\pm$0.03. The resulting fill-out factor is rather
hight, $F$ = 0.63$\pm$0.05. The best fits to the data are shown in 
Fig.~\ref{fig04}. The major deficiency of the fits is insufficient interpretation
of the primary minimum, which is sharper than predicted suggesting either higher
mass ratio, surface inhomogeneities or deficiency of the Roche model for this system.
It cannot be excluded that the observed LC is affected by some night-to-night changes
resulting in systematically wrong parameters and fits.

\section{Long-term evolution of orbits}
\label{secular}

The ratio of the orbital periods of the outer orbit and orbit of the non eclipsing
pair is $P_{1234}/P_{34}$ = 44.76. Corresponding period ratio in the case of the 
contact binary is as large as $P_{1234}/P_{12}$ = 743. Therefore, visible 
precession of the orbit can be detected (on human timescales) only in the case 
of the detached pair.

According to \citet{sode1975} the nodal precession period caused by the perturbations
in the close triple is:

\begin{equation}
P_{node} = \frac{4}{3} \left(\frac{M_{1234}}{M_{34}} \right) 
\frac{P^2_{1234}}{P_{34}} (1-e_{1234}^2)^{3/2} \frac{L_{1234}}{(L_{34} + L_{1234}) \cos j}
\end{equation}

where $L_{34}$ is angular momentum in the detached binary orbit, 
and $L_{1234}$ is (vector) sum of angular momenta in the 
detached binary orbit and the outer orbit. 
In our case we can neglect the term in the last parentheses since 
$(L_{34} + L_{1234})/L_{1234} \sim 1.08$ and orbits are close to being coplanar; 
hence $\cos j \sim 1.00$. Then we get $P_{node} \sim = 120$ years. Since the detached pair 
on $P_{34}$ = 7.93-days orbit is non-eclipsing binary 
(at least at present) we can detect possible precession of the orbit only from the RVs.
Precession of the orbital plane should results in changes of the amplitude of RVs
and apsidal motion.

Possibility of long-term changes of the 7.93-days orbit was assessed by dividing 
RV observations into individual observing seasons. Since LITE in the 
eclipsing pair is not affected nor RVs of its components, simultaneous fits in 
separate time intervals used all measurements for the eclipsing pair with the parameters
of the contact-binary orbit and outer orbit fixed to those of the global solution.

Five solutions given in Table~\ref{tab06} can be described/interpreted as follows: 
(i) eccentricity is stable - solutions are reliable and small eccentricity is not spurious
(ii) semi-amplitudes of RVs remain stable within the errors - mutual orbit and orbit of the
detached pair are very probably close to coplanar (iii) there is definite apsidal motion.
The coplanarity of the orbits means that the orbit of the non-eclipsing binary very 
probably not changes its inclination angle and we cannot expect appearance of the 
eclipses in this binary in future (see \citet{may2005}). The period of the line of apsides motion can roughly
be estimated from longitude of periastron in 1998 and in 2007/2008 season as about 80 years.
Further observations are, however, needed to precisely determine the rate of the
apsidal advance and its quantitative interpretation in terms of the tidal distortion, 
relativistic and gravitational perturbation effects.

\begin{table*}
\caption{Long-term evolution of the non-eclipsing binary orbit. Parameters of
         contact-binary orbit and wide mutual orbit were adopted from the 
         global solution in \ref{tab05}. Standard error of the last digit is
         given in parenthesis. \label{tab06}}
\footnotesize
\begin{center}
\begin{tabular}{lccccc}
\hline
\hline
Parameter            &     1998    &  1999-2000  &     2002    &  2004-5     &  2007-8     \\
HJD (2\,400\,000+)   & 50852-50960 & 51261-51673 & 52277-52391 & 53060-53836 & 53823-54650 \\
\# of points         &     40      &     19      &    73       &   71        &    36       \\
$e_{34}$             &   0.032(6)  &   0.030(6)  &  0.032(5)   &  0.038(4)   &  0.039(4)   \\
$\omega_{34}$ [rad]  &   1.43(19)  &   1.81(12)  &   2.04(15)  &  2.17(14)   &  2.18(9)    \\
$K_3$ [km~s$^{-1}$]  &   63.69(37) &  62.72(27)  &  65.00(39)  & 63.72(28)   &  63.59(34)  \\
$K_4$ [km~s$^{-1}$]  &   65.67(42) &  66.47(32)  &  65.51(43)  & 65.22(33)   &  65.79(39)  \\
\hline
\hline
\end{tabular}
\end{center}
\end{table*}

\section{Possible astrometric and direct detection}

VW~LMi (HIP 54003) was astrometrically observed during the Hipparcos mission. No positional
disturbance or acceleration terms in the proper motion were found and there is no
Double/Multiple Systems Annex flag (H59 field) in spite of the fact that VW~LMi
is member of multiple system with orbital period much shorter than the time span of the
Hipparcos astrometry. The apparent simplicity of VW~LMi is reflected in small formal
error of its annual parallax, 0.90 mas.

The reasons for negative detection in the Hipparcos astrometry are very probably two
(i) the orbital period is close to one year and orbital wobble mimics parallactic motion
and changes apparent annual parallax (ii) Hipparcos astrometry refers to the photocenter 
of VW~LMi which substantially decreases the astrometric wobble. Ecliptical latitude 
of VW~LMi is about $\beta$ = 22.325\degr. The inclination angle of the outer orbit 
is $i_{1234} = 64.1\pm$4.2\degr (see Section~\ref{absolute}). Then $\sin \beta \approx 
\cos i_{1234}$, which means that parallactic ellipse and relative orbit of the 
photocenter would be ellipse of similar shape (with practically the same period of 
apparent motion). The chances of disentangling the effects and solving the problem 
depend on the longitude of the ascending node. 

Outer-orbit parameters (Table~\ref{tab05}) provide projected major axis as 
$a_{1234} \sin i_{1234}$ = 1.456$\pm$0.019 AU. Neglecting small eccentricity
of the orbit the maximum angular separation of the components is $\pi a_{1234}$.
Assuming $i_{1234} = 64.1\pm4.2 \degr$ (Section \ref{absolute}) the maximum angular
separation is just about 13 mas.

The full amplitude of the photocenter motion can be found assuming 
$(L_3 + L_4)/(L_1 + L_2) = 0.42$ and $a_{34}/a_{12}$ = 1.09 and $\alpha$ = 13 mas as: 

\begin{equation}
a_{phot} = \alpha \frac{L_{34} a_{34} - L_{12} a_{12}}{L_{34} + L_{12}},
\end{equation}

For VW~LMi we get $a_{phot}$ = 5.0 mas. This is comparable to the typical residuals 
seen in the Hipparcos astrometric solutions. Some improvement of the astrometric solution 
can be obtained using proper motion as known parameter from ground-based observations.
The solving of the orbit might be complicated by variability induced motion (VIM) caused by 
light changes of the eclipsing pair. Therefore, VW~LMi is rather difficult target 
for astrometric modelling. With expected maximum separation of the components of about 
13 mas and with total brightness only $V_{max}$ = 8.08 ($K$ = 7.20) it would not be easy 
target neither for interferometric observations. 

\section{Absolute parameters of the components}
\label{absolute}

Parameters of the components can be determined by a simple procedure used in
\citet{ddo11}. Using new determination of the inclination angle $i$ = 79.0\degr~
and the projected total mass of the contact pair $(M_1+M_2) \sin^3 i_{12} =
2.231\pm0.023$ M$_\odot$, we obtain $(M_1 + M_2) = 2.359\pm0.024$ M$_\odot$
\footnote{we neglect error of the inclination angle depending on spectroscopic
          determination of the third light which can  systematically be wrong}. 
The outer, 355-days orbit defines the mass ratio for the two pairs,
$(M_1+M_2)/(M_3+M_4) = 1.074\pm0.025$. Therefore, the true (not the
projected) mass of the second spectroscopic binary is $(M_3+M_4) =
2.196\pm0.073 M_\odot$. Using the projected mass $(M_3+M_4) \sin^3 i_{34} =
1.785\pm0.011$ M$_\odot$, we estimate the inclination of the orbit
of the second pair to be about $68.9\pm1.5 \degr$. The outer, 355-day
orbit is even less inclined to the sky: with
$(M_1+M_2+M_3+M_4) = 4.56\pm$0.07 M$_\odot$ and the projected total mass
of only 3.32$\pm$0.10 M$_\odot$, we obtain $i_{1234}$ = 64.1$\pm$4.2$\degr$.
In the view of large uncertainties, the planes of the detached pair 
and mutual wide orbit could still be coplanar. 

Individual masses of components are
then $M_1$ = 1.66 M$_\odot$, $M_2$ = 0.70 M$_{\odot}$, $M_3$ = 1.11 M$_{\odot}$, and 
$M_4$ = 1.09 M$_{\odot}$. Masses of the components are rather inconsistent with
the fact that all four components show practically the same spectral type. 
Assuming $L \sim M^{3.45}$ and that all four components produce energy as MS
stars results in $L_{34}/L_{12}$ = 0.49. In the region of Mg I triplet we arrived at 
rather compatible value. The only remaining problem is the fact that the third light
as found from photometry is virtually the same in the $B$ and $V$ passbands 
indicating the same $T_{eff}$ for both binaries which is incompatible with their masses.
The second binary should have components of about G0V spectral type according to their
masses. Due to the fact that we have spectra just in short wavelength region,
the individual component spectra cannot be reliably disentangled. The issue could
be resolved either by echelle spectroscopy and/or multi-color photometry. 

\section{Conclusions}

Tightest known quadruple system VW~LMi is really unique. It definitely deserves
additional observations and analysis. The study of this system could bring light
on (i) the evolution and origin of binary stars in multiple systems of stars 
(ii) tidal interaction of third body and its influence on the Roche geometry 
(iii) long-term evolution of orbits in tight multiple systems. The system is
useful for further analysis since all four components are visible in the spectrum.
The determination of the individual component's parameters like $T_{eff}$, $\log g$,
metalicity could benefit from spectra disentangling (see \citet{hadr1995}).
The absolute parameters of all components which had to originate at practically
same time could be used to determine the evolutionary status and history
of the system. 

The study of VW~LMi is, however, complicated by (i) the outer orbital period being close to
one year which complicates its phase coverage by Earth-bound observer (ii) very small
angular separation of the components making direct resolving of the 
mutual wide orbit and reliable determination of $i_{1234}$ difficult (which would definitely
improve determination of individual masses), (iii) fast rotation of the components 
of the contact binary making RV measurements unsure and rectification to real continuum 
impossible (due to the line-blanketing) (iv) eclipses and Roche geometry in the 
contact pair making usual assumption of the spectra disentangling techniques not
valid (line profiles of individual components cannot change with orbital revolution). 

It is also interesting to note, that VW~LMi and HD95606 show very similar proper 
motion, parallax and RVs (see Table~\ref{tab02}) - the stars definitely form 
a loosely bound pair and all components very probably evolved from the same 
protostellar cloud (see \citet{oswa2007}).

The orbit of the detached pair in VW~LMi with 7.93-days period is almost circular. 
This is the case of another two quadruples detected by the DDO observations, 
TZ~Boo and V2610~Oph (DDO series No. XIV, \citet{ddo14}). That means that either 
it evolved with such orbit or it was circularized by the gravitational interaction 
with the contact binary. Detected apsidal motion in VW~LMi requires further observing 
to reliably determine apsidal period. Investigation of the observed eccentricities
of the second binaries and their predicted synchronization timescales in quadruple 
systems with contact binaries could shed light on the age and evolutionary status
of contact binaries \citep{ruci-priv2008}.

Better characterization of VW~LMi calls for long-term monitoring to cover
the whole 355-days orbital cycle. Especially, times of minima should be obtained
free of any systematic effects. The understanding of the system would greatly
benefit from visual orbit obtained by means of long-baseline interferometry.
Multi-color photometry and/or echelle spectroscopy could lead to reliable determination
of component's temperatures and luminosities.

\medskip

The stays of TP at DDO have been supported by a grant to Slavek M. Rucinski from
the Natural Sciences and Engineering Council of Canada.
This research has been supported in part by the Slovak
Academy of Sciences under grants No. 2/7010/7 and 2/7011/7, and
grant of the \v{S}af\'arik University VVGS 9/07-08.
MV's research is supported by a Marie Curie ``Transfer of Knowledge''
Fellowship within the 6th European Community Framework Programme.
The observations at Astronomical Observatory at Kolonica Saddle 
(part of Vihorlat Observatory) were partially supported by APVV grant 
LPP-0049-06 and APVV bilateral grant SK-UK-01006.
DB thanks to Miron Kerul-Kmec for technical assistance with CCD camera
at the Roztoky Observatory.

\label{lastpage}

\end{document}